\documentclass[prl,twocolumn,showpacs,preprintnumbers,amsmath,amssymb,article,superscriptaddress,svgnames]{revtex4-1}
\pdfoutput=1
\usepackage{graphicx}
\usepackage{float}
\usepackage{verbatim}
\usepackage{dcolumn}
\usepackage{dsfont}
\usepackage{color}
\usepackage[utf8]{inputenc}
\usepackage{amsmath}
\usepackage[makeroom]{cancel}
\usepackage{amsfonts}
\usepackage{amssymb}
\usepackage{xcolor}
\usepackage{xspace}
\usepackage{physics}
\usepackage[mathscr]{euscript}
\usepackage{stix}
\usepackage{siunitx}
\usepackage{bm}
\usepackage[colorlinks=true, linkcolor=blue, citecolor=purple, urlcolor=blue]{hyperref}
\usepackage{makecell}

\newcommand\nm[1]{\SI{#1}{\nano\meter}}

\begin{document}


\newcommand{\beq}{\begin{equation}}
\newcommand{\eeq}{\end{equation}}
\newcommand{\beqa}{\begin{eqnarray}}
\newcommand{\eeqa}{\end{eqnarray}}
\newcommand{\lf}{\hfil \break \break}
\newcommand{\ahat}{\hat{a}}
\newcommand{\adag}{\hat{a}^{\dagger}}
\newcommand{\adagg}{\hat{a}_g^{\dagger}}
\newcommand{\bhat}{\hat{b}}
\newcommand{\bdag}{\hat{b}^{\dagger}}
\newcommand{\bdagg}{\hat{b}_g^{\dagger}}
\newcommand{\chat}{\hat{c}}
\newcommand{\cdag}{\hat{c}^{\dagger}}
\newcommand{\dhat}{\hat{d}}
\newcommand{\nhat}{\hat{n}}
\newcommand{\ndag}{\hat{n}^{\dagger}}
\newcommand{\den}{\hat{\rho}}
\newcommand{\phihat}{\hat{\phi}}
\newcommand{\Ahat}{\hat{A}}
\newcommand{\Adag}{\hat{A}^{\dagger}}
\newcommand{\Bhat}{\hat{B}}
\newcommand{\Bdag}{\hat{B}^{\dagger}}
\newcommand{\Chat}{\hat{C}}
\newcommand{\Dhat}{\hat{D}}
\newcommand{\Ehat}{\hat{E}}
\newcommand{\Lhat}{\hat{L}}
\newcommand{\Nhat}{\hat{N}}
\newcommand{\Ohat}{\hat{O}}
\newcommand{\Odag}{\hat{O}^{\dagger}}
\newcommand{\Shat}{\hat{S}}
\newcommand{\Uhat}{\hat{U}}
\newcommand{\Udag}{\hat{U}^{\dagger}}
\newcommand{\Xhat}{\hat{X}}
\newcommand{\Zhat}{\hat{Z}}
\newcommand{\Xdag}{\hat{X}^{\dagger}}
\newcommand{\Ydag}{\hat{Y}^{\dagger}}
\newcommand{\Zdag}{\hat{Z}^{\dagger}}
\newcommand{\Ham}{\hat{H}}
\newcommand{\bis}{{\prime \prime}}
\newcommand{\tris}{{\prime \prime \prime}}
\newcommand{\bracket}[3]{\mbox{$\langle#1|#2|#3\rangle$}}
\newcommand{\mat}[1]{\overline{\overline{#1}}}
\newcommand{\dotp}{\mbox{\boldmath $\cdot$}}
\newcommand{\tp}{\otimes}
\newcommand{\hak}[1]{\left[ #1 \right]}
\newcommand{\vin}[1]{\langle #1 \rangle}
\newcommand{\tes}[1]{\left( #1 \right)}
\newcommand{\nav}{\langle \hat{n} \rangle}

\newcommand{\macro}[1]{\texttt{\textbackslash#1}}
\newcommand{\env}[1]{\texttt{#1}}
\newcommand{\ud}[1]{{#1^{\dagger}}}
\newcommand{\av}[1]{\langle  #1 \rangle}
\newcommand{\pop}[1]{\langle \ud{#1} #1 \rangle}
\newcommand{\corr}[3]{\langle #1^{\dagger #2} #1^{#3} \rangle}

\providecommand{\tabularnewline}{\\} 
\newcommand{\mean}[1]{\langle#1\rangle}
\newcommand{\bb}[1]{\mathbb{#1}}
\newcommand{\cc}[1]{#1^{\ast}}

\hyphenation{Teich}

\title{The Origin of Antibunching in Resonance Fluorescence}

\author{Lukas~Hanschke}
\thanks{L. H., L. S. and J. C. L. C. contributed equally to this work}
\affiliation{Walter Schottky Institut and Department of Electrical and Computer Engineering, Technische Universit\"at M\"unchen, 85748, Garching, Germany}
\affiliation{Munich Center of Quantum Science and Technology (MCQST), 80799 Munich, Germany}
\author{Lucas~Schweickert}
\thanks{L. H., L. S. and J. C. L. C. contributed equally to this work}
\affiliation{Department of Applied Physics, Royal Institute of Technology, Albanova University Centre, Roslagstullsbacken 21, 106 91 Stockholm, Sweden}
\author{Juan~Camilo~L\'opez~Carre\~{n}o}
\thanks{L. H., L. S. and J. C. L. C. contributed equally to this work}
\affiliation{Faculty of Science and Engineering, University of Wolverhampton,Wulfruna St, Wolverhampton WV1 1LY, United Kingdom}
\author{Eva~Sch\"oll}
\author{Katharina~D.~Zeuner}
\author{Thomas~Lettner}
\affiliation{Department of Applied Physics, Royal Institute of Technology, Albanova University Centre, Roslagstullsbacken 21, 106 91 Stockholm, Sweden}%
\author{Eduardo Zubizarreta Casalengua}
\affiliation{Faculty of Science and Engineering, University of Wolverhampton,Wulfruna St, Wolverhampton WV1 1LY, United Kingdom}
\author{Marcus~Reindl}
\author{Saimon~Filipe~Covre~da~Silva} 
\affiliation{Institute of Semiconductor and Solid State Physics, Johannes Kepler University Linz, 4040, Austria}
\author{Rinaldo~Trotta}
\affiliation{Dipartimento di Fisica, Sapienza Universit\`a di Roma, Piazzale A. Moro 1, I-00185 Roma, Italy}
\author{Jonathan~J.~Finley}
\affiliation{Munich Center of Quantum Science and Technology (MCQST), 80799 Munich, Germany}
\affiliation{Walter Schottky Institut and Physik Department, Technische Universit\"at M\"unchen, 85748, Garching, Germany}
\author{Armando~Rastelli}
\affiliation{Institute of Semiconductor and Solid State Physics, Johannes Kepler University Linz, 4040, Austria}
\author{Elena~del~Valle}
\affiliation{Faculty of Science and Engineering, University of Wolverhampton,Wulfruna St, Wolverhampton WV1 1LY, United Kingdom}
\affiliation{Departamento de F\'{i}sica T\'{e}orica de la Materia Condensada, Universidad Aut\'{o}noma de Madrid, 28049 Madrid, Spain}
\author{Fabrice~P.~Laussy}
\affiliation{Faculty of Science and Engineering, University of Wolverhampton,Wulfruna St, Wolverhampton WV1 1LY, United Kingdom}
\affiliation{Russian Quantum Center, Novaya 100, 143025 Skolkovo, Moscow Region, Russia}
\author{Val~Zwiller}
\affiliation{Department of Applied Physics, Royal Institute of Technology, Albanova University Centre, Roslagstullsbacken 21, 106 91 Stockholm, Sweden}
\author{Kai~M\"uller} 
\affiliation{Walter Schottky Institut and Department of Electrical and Computer Engineering, Technische Universit\"at M\"unchen, 85748, Garching, Germany}
\affiliation{Munich Center of Quantum Science and Technology (MCQST), 80799 Munich, Germany}
\author{Klaus~D.~J\"ons}
\email{corresponding author: klausj@kth.se}
\affiliation{Department of Applied Physics, Royal Institute of Technology, Albanova University Centre, Roslagstullsbacken 21, 106 91 Stockholm, Sweden}

\date{\today}

\begin{abstract}
Epitaxial quantum dots have emerged as one of the best single-photon sources, not only for applications in photonic quantum technologies but also for testing fundamental properties of quantum optics. One intriguing observation in this area is the scattering of photons with subnatural linewidth from a two-level system under resonant continuous wave excitation. In particular, an open question is whether these subnatural linewidth photons exhibit simultaneously antibunching as an evidence of single-photon emission. Here, we demonstrate that this simultaneous observation of subnatural linewidth and antibunching is not possible with simple resonant excitation. First, we independently confirm single-photon character and subnatural linewidth by demonstrating antibunching in a Hanbury Brown and Twiss type setup and using high-resolution spectroscopy, respectively. However, when filtering the coherently scattered photons with filter bandwidths on the order of the homogeneous linewidth of the excited state of the two-level system, the antibunching dip vanishes in the correlation measurement. Our experimental work is consistent with recent theoretical findings, that explain antibunching from photon-interferences between the coherent scattering and a weak incoherent signal in a skewed squeezed state. 

\end{abstract}

\maketitle

Quantum dots are ideally suited as prototypical two-level quantum systems in the solid state. This is a result of their strong optical interband transitions, almost exclusive emission into the zero-phonon line and ease of integration into opto-electronic devices~\cite{Trivedi2020, Borri2001, Brash2019, Senellart2017}. Moreover, the development of resonant excitation techniques~\cite{Muller2007}, such as cross-polarized resonance fluorescence~\cite{Kuhlmann2013} has enabled nearly transform-limited linewidths~\cite{Kuhlmann2015}, as the resonant excitation avoids the generation of free charge carriers which can lead to a fluctuating electronic environment resulting in spectral diffusion~\cite{Chen2016}. This technique has enabled multiple exciting tests of quantum optics as well as the use of quantum dots as sources of non-classical light for photonic quantum technologies~\cite{Senellart2017}. For example, using pulsed excitation, Rabi oscillations have been demonstrated and enabled the on-demand generation of single photons~\cite{He2013}, entangled photon pairs~\cite{Muller2014}, two-photon pulses~\cite{Fischer2017}, and photon number superposition states~\cite{Loredo2019}. Furthermore, continuous wave excitation has led to the observation of Mollow triplets for strong driving~\cite{Flagg2009} as well as coherent Rayleigh scattering in the regime of weak driving~\cite{Nguyen2011,Matthiesen2012,Konthasinghe2012}. In the latter case, light is coherently scattered by the two-level system leading to a subnatural linewidth of the photons which inherit the coherence of the laser~\cite{Matthiesen2013}. While previous experimental works have indicated that the coherently scattered light exhibits antibunching~\cite{Nguyen2011, Matthiesen2012}, recent theoretical studies have predicted that the antibunching is only enabled by the presence of weak incoherent emission interfering with the coherently scattered light~\cite{LopezCarreno2018}. Therefore, it was predicted that selectively transmitting the narrow coherent scattering by frequency filtering, i.e., suppressing the incoherently scattered component, would inhibit the observation of antibunching. In this letter, we experimentally test this prediction and observe that, indeed, it is only possible to observe either subnatural linewidth or antibunching under simple resonant excitation. We provide a fundamental theoretical model giving insight to the underlying mechanism which agrees very well with our experimental results without data processing. The excellent accord between experiment and theory indicates that targeted experiments to control the balance of coherent and incoherent fractions and simultaneously achieve antibunching and subnatural linewidth, are within sight.

\begin{figure}
    \centering
    \includegraphics[width=\columnwidth]{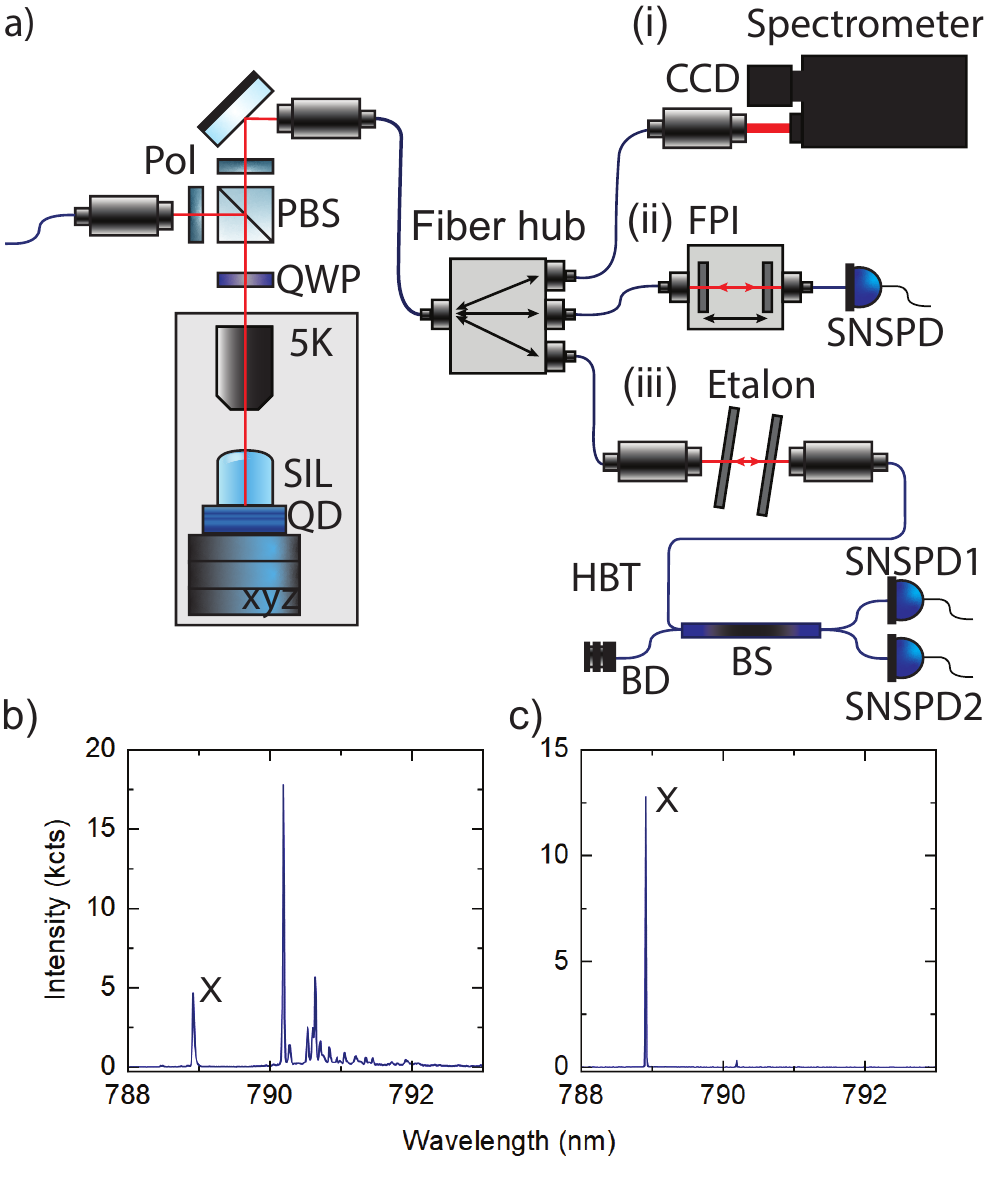}
    \caption{a) Experimental setup to generate coherently scattered photons from our GaAs quantum dot. Cross polarization using a polarizing beam splitter (PBS), nano particle polarizers (Pol) and a quarter waveplate (QWP). The photons scattered from the quantum dot are additionally spatially filtered from the excitation laser using a single-mode fiber. The quantum dot is located in a closed-cycle cryostat at $\SI{5}{\kelvin}$ temperature. A solid immersion lens (SIL) increases the collection efficiency of the emitted quantum light. The collected signal can be analyzed (i) in a spectrometer equipped with a silicon CCD, (ii) using a tunable Fabry-P\'erot interferometer (FPI) equipped with a superconducting nanowire single-photon detector (SNSPD), or (iii) by sending it through different types of frequency filters (Etalon) and then into a Hanbury Brown and Twiss (HBT) setup to measure the second-order intensity autocorrelation of the signal (BD = beam dump, BS = 50/50 beam splitter). b) Quasi-resonant excitation spectrum of the investigated quantum dot, using \nm{781} wavelength pulsed excitation. c) Resonance fluorescence spectrum of the same quantum dot as graph b). The exciton (X) is excited resonantly with a narrow-band continues wave diode laser.} 
    \label{fig:setup}
\end{figure}

The quantum dots used in this study were grown by droplet etch epitaxy~\cite{Heyn2009, Huo2013}. An aluminum droplet is used to dissolve an AlGaAs substrate at distinct positions to form near perfectly round holes with a diameter of $\sim$$\SI{100}{\nano\meter}$ and $\sim$$\SI{5}{\nano\meter}$ depth. These holes are filled with GaAs in a second step and capped again by AlGaAs to form single quantum dots. A frequency tunable diode laser with a narrow linewidth of $\SI{50}{\kilo\hertz}$ is used to resonantly excite a single quantum dot. To suppress the leakage of laser light into the detection path of the setup we use a pair of perpendicular thin film polarizers in the excitation and detection paths. The emitted photons are further filtered with a self-build transmission spectrometer with a FWHM of $\SI{19}{\giga\hertz}$ to suppress any residual emission of other transitions. Figure~\ref{fig:setup}\,a) depicts the setup used for this experiment which can be used either to introduce a Hanbury Brown and Twiss setup to investigate the photon statistics or a scanning Fabry-P\'erot cavity to obtain high resolution spectra. By populating higher excited states of the quantum dot with a laser that is at the same time mixed with low intensity white light to stabilize the electrical environment \cite{Nguyen2012} we obtain the spectrum shown in Fig.~\ref{fig:setup}\,b). Several emission lines appear, among which we can identify the neutral exciton transition. Switching to resonant excitation leads to a clean spectrum with only a single peak from the excited transition, shown in Fig.~\ref{fig:setup}\,c).

\begin{figure}
    \centering
    \includegraphics[width=\linewidth]{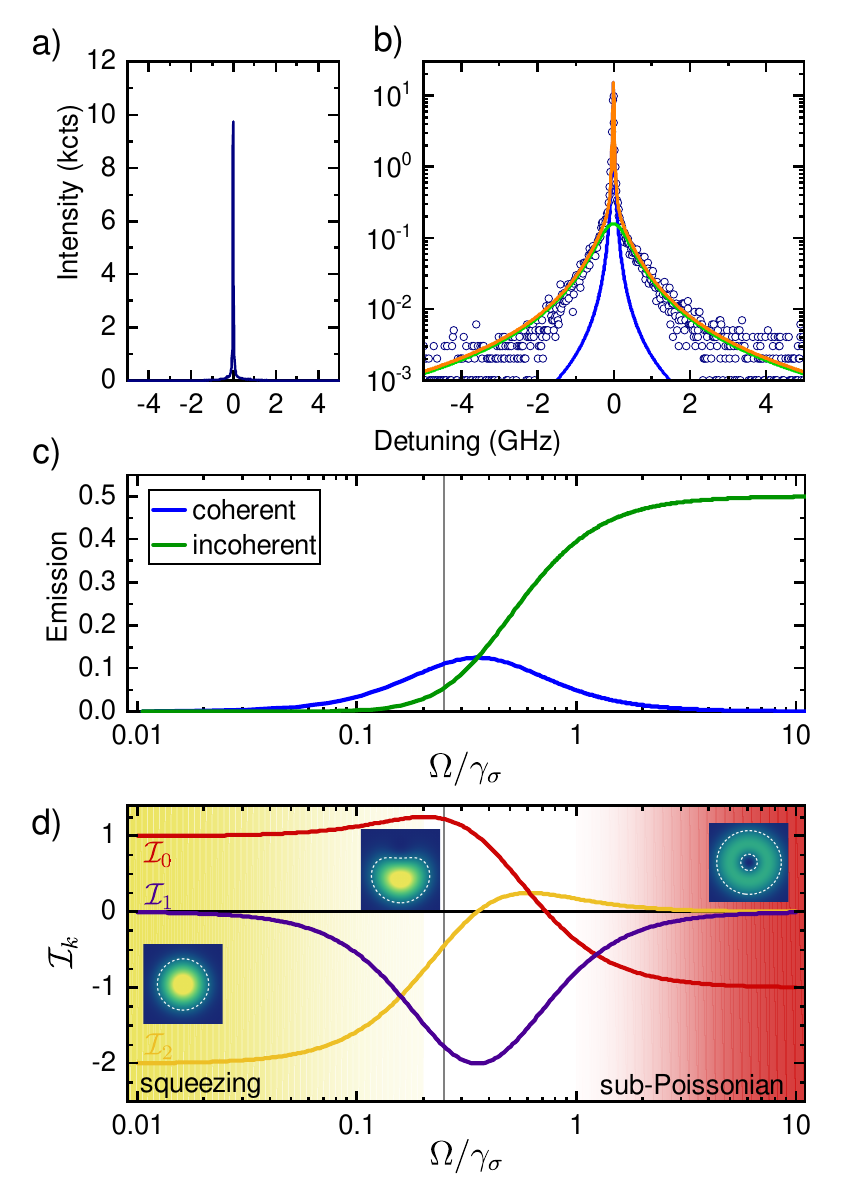}
    \caption{a) High resolution spectrum of the exciton transition under resonant excitation in the weak pumping regime.
    b) The spectrum plotted in semi-logarithmic scale reveals a second broader peak. Blue line: coherently scattered laser, green line: incoherent resonance fluorescence; orange line: cumulative peak. c) Theoretical curve of the intensity of the coherent and incoherent component as a function of the driving power. d)  Two-photon interference terms~$\mathcal{I}_k$,
    Eqs.~(\ref{eq:Thu30Apr193821CEST2020}), with~$k=0,2$ playing a
    role at weak and strong drivings and showing how
    antibunching~$g^{(2)}(0)=0$ arises from squeezing
    (with~$\mathcal{I}_2=-2$ on the left) or sub-Poissonian statistics of the
    emitter (with~$\mathcal{I}_0=-1$, on the right). The transition
    between the two regimes occurs through a skewing of the squeezed
    state whereby $\mathcal{I}_2$ gets replaced by
    $\mathcal{I}_1$. Insets: the Wigner representation $W_\sigma(X,Y)$ of the quantum state at weak, intermediate and strong driving, for $-1.5\le X,Y,\le 1.5$ with white dashed isolines at 0 and 0.1. Note that at strong driving, $W_\sigma$ becomes negative (non-Gaussian). The vertical line indicates the driving of our
    experiment.}
    \label{fig:FPI}
\end{figure}

We now focus on studying the emission under resonant excitation using a scanning Fabry-P\'erot interferometer with a spectral resolution of $\SI{28}{\mega\hertz}$. While in a linear scale (Fig.~\ref{fig:FPI}\,a)) the spectrum seems to consist of only one sharp peak, a plot in logarithmic scale (Fig.~\ref{fig:FPI}\,b)) reveals the presence of two superimposed peaks: A sharp peak with a linewidth of $\SI{28}{\mega\hertz}$ and a broader peak with a linewidth of $\SI[separate-uncertainty = true]{890(60)}{\mega\hertz}$. While the sharp peak stems from the coherent scattering and is only limited by the resolution of the scanning Fabry-P\'erot interferometer, the broader peak stems from incoherent emission. Here, the observed linewidth results from emission mainly given by the Fourier-limit. The ratio of the integrated peak areas is 1:2.65 and consistent with the numerical simulation of a resonantly driven two-level system (Fig.~\ref{fig:FPI}\,c)) where for weak driving the coherent scattering dominates while for strong driving the situation is reversed.

\begin{figure}
    \centering
    \includegraphics[width=\linewidth]{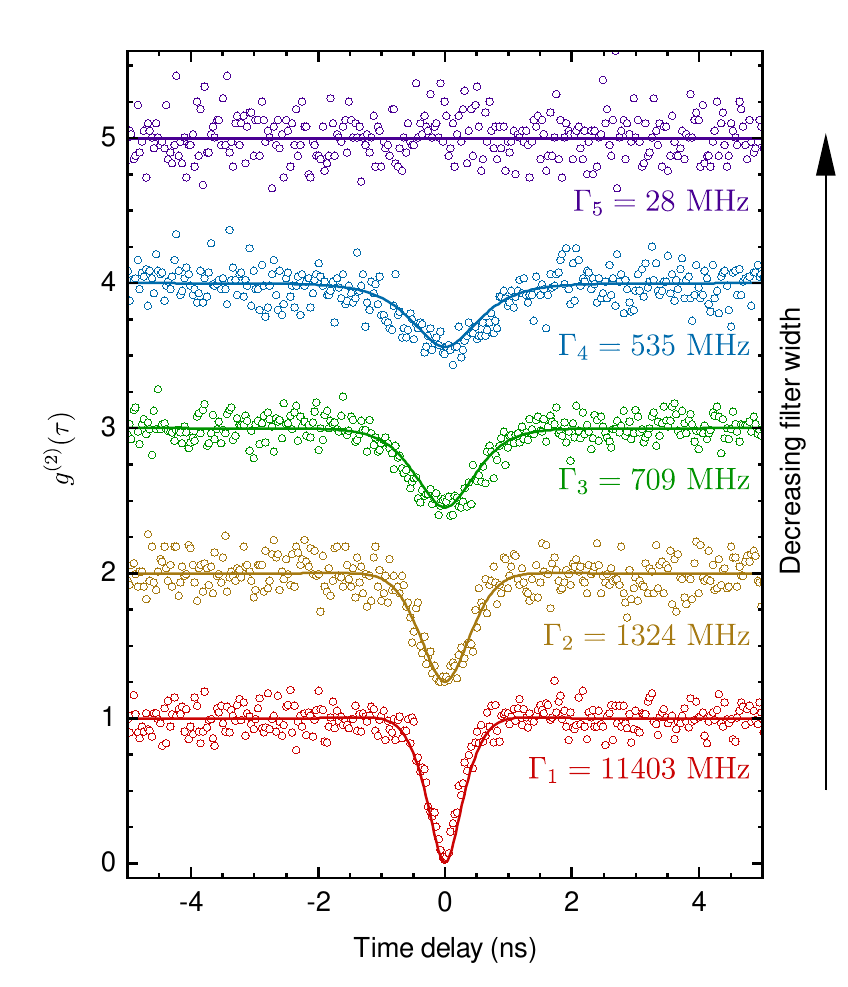}
    \caption{Second-order intensity correlation function~$g^{(2)}(\tau)$
    of the quantum dot emission in the Rayleigh regime for different
    spectral filter widths $\Gamma_x$. With decreasing filter width, a larger
    portion of the incoherent component is suppressed, unbalancing the two-photon interference which
    produces antibunching in this regime.  The experimental data is
    shown with empty circles, while the solid lines are obtained with
    the theory of frequency-resolved correlations using the parameters
    given in Table~\ref{tab:Fri1May111809CEST2020}. 
    }
    \label{fig:g2}
\end{figure}

To verify the single-photon character of the quantum dot emission, we perform second order intensity autocorrelation measurements using a Hanbury Brown and Twiss setup connected to two superconducting nanowire single-photon detectors, with low dark count rates~\cite{Schweickert2018}. Our Hanbury Brown and Twiss setup has a time resolution of $\SI{70}{\pico\second}$ given by the internal response function. The unfiltered emission in the Rayleigh regime shows near perfect antibunching Fig.~\ref{fig:g2} (red), confirming the single-photon character, with a measured degree of second-order coherence of $g^{(2)}(0)=0.022 \pm{0.011}$. For this measurement we used a broad frequency filter of FWHM = $\SI{19}{\giga\hertz}$, more than 20 times broader than the linewidth of the incoherent emission.

The light emitted by an ideal two-level system under perfect detection
conditions is always antibunched, but the physical mechanism for this
depends on the regime in which it is being excited. In the case of
coherent driving by a laser, one can distinguish between the
weak-driving Rayleigh regime where antibunching is due to a
coherent process of absorption and re-emission of the incident
coherent radiation by the two-level system~\cite{Heitler1954}, and
the strong-driving limit, where the two-level system blocks the
excitation, gets saturated and emits antibunched light in the fashion
of the spontaneous emission of a two-level system. While one has in
mind the second mechanism when thinking of antibunching from a
two-level system, the first mechanism is completely unrelated and must
be understood instead as an interference
effect~\cite{Casalengua2020}. The two-level system annihilation
operator~$\sigma$ can be decomposed into a sum of a coherent
term~$\langle\sigma\rangle$ and a quantum, or incoherent,
term~$\varsigma\equiv\sigma-\langle\sigma\rangle$ as:
\begin{equation}
  \sigma=\langle\sigma\rangle+\varsigma\,.
\end{equation}
Note that~$\varsigma$ is an operator, like~$\sigma$, in fact it is
simply $\sigma$ minus its coherent part~$\langle\sigma\rangle$. Their
respective intensities as a function of the driving~$\Omega$ and emission rate $\gamma_{\sigma}$ are given
by~\cite{Meystre2007}:
\begin{equation}
  \label{eq:Thu30Apr194924CEST2020}
  |\langle\sigma\rangle|^2=\frac{4\gamma_\sigma^2\Omega^2}{(\gamma_\sigma^2+8\Omega^2)^2}\quad\text{and}\quad
  \langle\varsigma^\dagger\varsigma\rangle=\frac{32\Omega^4}{(\gamma_\sigma^2+8\Omega^2)^2}
\end{equation}
and are shown in Fig.~\ref{fig:FPI}\,(c). While the
total intensity $n_\sigma\equiv\langle\sigma^\dagger\sigma\rangle$ for
the sum of these two fields would typically involve an interference
term
$n_\sigma=|\langle\sigma\rangle|^2+\langle\varsigma^\dagger\varsigma\rangle+2\mathrm{Re}\big(\langle\sigma\rangle^*\langle\varsigma\rangle\big)$,
in this case there is no interference
since~$\langle\varsigma\rangle=0$ by construction ($\varsigma$ has no
mean field). Higher-order photon correlations, however, do exhibit
such interferences between the coherent component
$\langle\sigma\rangle$, which inherits the statistics of the laser, and
$\varsigma$, which follows the statistics of the two-level system's
quantum fluctuations. Such interferences, at the two-photon level, are
quantified by coefficients~$\mathcal{I}_k$ which add up to the
zero-delay two-photon coherence function~$g^{(2)}(0)$ as
follows~\cite{Mandel1982,Carmichael1985,Vogel1991}:
\begin{equation}
  \label{eq:Fri1May091452CEST2020}
  g^{(2)}(0)=1+\mathcal{I}_0+\mathcal{I}_1+\mathcal{I}_2  ,
\end{equation}
where:
\begin{subequations}
  \label{eq:Thu30Apr193821CEST2020}
\begin{align}
  \mathcal{I}_0 &= \frac{\corr{\varsigma}{2}{2} -
                  \pop{\varsigma}^2}{\mean{\sigma^\dagger\sigma}^2}\,,\\
  \mathcal{I}_1 &= 4\frac{\Re[\langle\sigma\rangle^{*}\av{\ud{\varsigma}\varsigma^2}]}{\mean{\sigma^\dagger\sigma}^2}\,,\\
  \mathcal{I}_2 &= \frac{\av{X_{\varsigma,\phi}^2}-\av{X_{\varsigma_,\phi}}^2}{\av{\sigma^\dagger\sigma}^2} ,
\end{align}
\end{subequations}
and~$X_{\varsigma,\phi}=(e^{i\phi}\varsigma^\dagger+e^{-i\phi}\varsigma)/2$
is the $\varsigma$-field quadrature. $\mathcal{I}_0$ describes the sub-Poissonian
(when negative) or super-Poissonian (when positive) character of the
quantum fluctuations, $\mathcal{I}_1$ its so-called anomalous
moments~\cite{Mandel1982,Vogel1991} and $\mathcal{I}_2$ its squeezing \cite{Loudon2000}
(when negative). These quantities are shown in
Fig.~\ref{fig:FPI}\,d), where one can see the
transition from $\mathcal{I}_0=1$ to~$-1$ when going from weak to
strong driving, which is compensated by the transition from
$\mathcal{I}_2=-2$ to~$0$ to keep the
total~(\ref{eq:Fri1May091452CEST2020}) zero.  To keep this identically
zero also in the transition between these two regimes, the system
develops a skewness in its squeezing through the anomalous
correlation term~$\mathcal{I}_1$ that overtakes~$\mathcal{I}_2$,
with~$\langle\ud{\varsigma}\varsigma^2\rangle$ becoming non-zero (it
cannot be factored into
$\langle\ud{\varsigma}\varsigma\rangle\langle\varsigma\rangle$
anymore), in such a way as to satisfy
$\mathcal{I}_1=-(1+\mathcal{I}_0+\mathcal{I}_2)$~\cite{Casalengua2020}. The
numerator $\Re [ \av{\sigma}^* \av{\varsigma^{\dagger} \varsigma^2} ]$
can be written as
$|\av{\sigma}| \big(\av{{:}X^3_{\varsigma, \phi}{:}} +
\av{{:}X_{\varsigma, \phi} Y^2_{\varsigma, \phi}{:}} \big)$
with~$Y_{\varsigma,\phi} \equiv \frac{i}{2} \left(e^{i \phi}
  \ud{\varsigma} - e^{-i \phi} \varsigma \right)$ the other $\varsigma$-quadrature and~$::$ denoting normal-ordering of the enclosed
quantity. This shows that, at weak driving, $\mathcal{I}_1$ becomes nonzero when the quantum state departs from a Gaussian description (squeezed thermal state) in the transition to the strong driving regime where it acquires the full non-Gaussian character of a single-photon source that is produced by a Fock state. Indeed, the full emission at strong driving comes exclusively from the quantum part $\sigma\approx\varsigma$, with the
system getting into the statistical mixture
$\rho=\frac{1}{2}(|0\rangle\langle0|+|1\rangle\langle1|)$, with no
coherence involved, $\langle\sigma\rangle=0$. Accordingly, the
sub-Poissonian statistics reaches its minimum~$\mathcal{I}_0=-1$.  In
the weak driving regime, antibunching is, on the opposite, due to
squeezing of the quantum fluctuations~$\varsigma$, with the system
being in a pure or skewed squeezed thermal state, with either
$\mathcal{I}_2$ or~$\mathcal{I}_1$ being $-2$, interfering with the
coherent component~$\langle\sigma\rangle$ to produce~$g^{(2)}(0)=0$.
This is even more clearly illustrated by considering the Wigner representation of the quantum state, as shown by the insets in Fig. \ref{fig:FPI} d) in the three regimes of interest, where one can see how the system evolves from a Gaussian state (a displaced squeezed thermal state) to a Fock state (a ring with a distribution that admits negative values) passing by a skewed (bean-shaped) Wigner distribution at the point of our experiment. Note that in the weak-driving regime, both the displacement and the ellipticity of the displaced squeezed thermal state are too small to be seen compared to the dominant thermal distribution, but both are necessary to produce antibunching. Counter-intuitively, at weak and intermediate driving, in direct opposition to the strong-driving case, quantum fluctuations are in fact
super-Poissonian, with~$\mathcal{I}_0\ge 1$. It is the interference
between such superbunched quantum fluctuations with the coherence of
the mean-field that result in an overall antibunching, this being the
two-photon counterpart of the apparent paradox of two waves adding to
produce no signal (destructive interferences).  
This understanding of the nature of antibunching in the Rayleigh regime is
important because attributing the non-Gaussian antibunching to the
scattered light makes it tempting to regard the scattered light as
having both the spectral feature of the laser, with a narrow
linewidth, and the statistical property of a two-level system,
antibunched. It has, indeed, been hailed as such in the
literature~\cite{Nguyen2011,Matthiesen2012}. As we have shown, however,
the Rayleigh antibunching does not come from the two-level character of the
emitter, which is not involved at such weak drivings, but from the
interference between the mean-field~$\langle\sigma\rangle$ as driven
by the laser (coherent absorption) and the quantum fluctuations
$\sigma-\langle\sigma\rangle$ (incoherent re-emission). Because it is
due to some interference, any tampering with the balance
$\mathcal{I}_0+\mathcal{I}_1+\mathcal{I}_2=-1$, for instance by frequency filtering,
will result in spoiling the antibunching~$g^{(2)}(0)=0$. Filtering is a
fundamental process in any quantum-optical measurement, since beyond
the finite bandwidth of any physical detector, a measurement that is
accurate in time requires detections at all frequencies and,
vice-versa, spectrally resolving emission requires integration over
time. To challenge the naive picture that light coherently-scattered
from a two-level system is antibunched, we measure $g^{(2)}(\tau)$ for
decreasing filter widths that increasingly isolate the coherent component. According to this naive picture, this should not affect
the property of light since the ``single photons'' are spectrally
sharp and will pass through the filter which does not block at their
frequency. According to the Rayleigh picture of interferences, however,
this will disrupt the balance of the~$\mathcal{I}_{k}$ coefficients in
their two-photon interference to produce antibunching.  The theory
shows that, for zero-delay coincidences in the weak-driving regime, the coefficients vary as a
function of filtering~$\Gamma$ as~\cite{Casalengua2020a}:
\begin{equation}
  \label{eq:Fri1May102246CEST2020}
  \mathcal{I}_0=\frac{\Gamma^2}{(\Gamma+\gamma_\sigma)^2}\,,\quad
  \mathcal{I}_1=0\,,\quad
  \mathcal{I}_2=-\frac{2\Gamma}{\Gamma+\gamma_\sigma}\,,
\end{equation}
with, therefore (cf.~Eq.~(\ref{eq:Fri1May091452CEST2020}))
\begin{equation}
  \label{eq:Fri17Apr143329CEST2020}
  g^{(2)}(0)=\left(\frac{\gamma_\sigma}{\Gamma+\gamma_\sigma}\right)^2\,.
\end{equation}
As these expressions show, filtering affects more the $\varsigma$
statistics than it does affect the squeezing of its quadratures.
This behaviour can be reproduced in the experiment by inserting a narrow spectral filter in the detection path. Measurements of $g^{(2)}(\tau)$ for different filter widths of $\SI[separate-uncertainty = true]{1550(320)}{\mega\hertz}, \SI[separate-uncertainty = true]{780(160)}{\mega\hertz}, \SI[separate-uncertainty = true]{390(80)}{\mega\hertz}$ and $\SI{28}{\mega\hertz}$ are presented as yellow, green, blue and purple data points in Fig.~\ref{fig:g2}, respectively. The data are offset in vertical direction for clarity. Clearly, with decreasing filter width, the depth of the antibunching dip decreases until it completely vanishes.

\begin{figure}
    \centering
    \includegraphics[width=\linewidth]{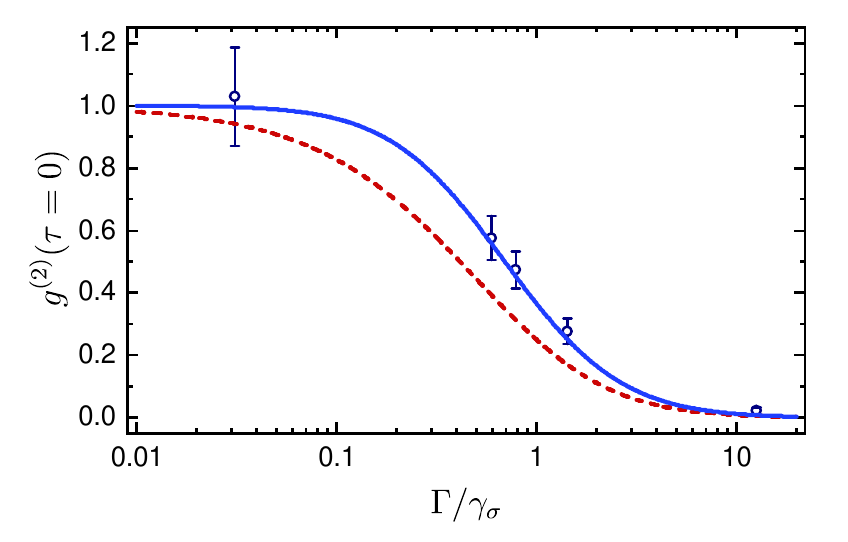}
    \caption{Loss of antibunching in the Rayleigh regime due to
    filtering~$\Gamma$. Dashed-red line, the limit of vanishing driving,
    Eq.~(\ref{eq:Fri17Apr143329CEST2020}), and solid-blue line, the case
    of small but finite driving~$\Omega$,
    Eq.~(\ref{eq:Fri1May105704CEST2020}). Our experimental data fits
    perfectly with the theoretical prediction.
    }
    \label{fig:theory}
\end{figure}

This is in excellent agreement with our theoretical model, that describes finite $\tau$-delay coincidences of the filtered light with an exact theory of time- and
frequency-resolved photon correlations~\cite{DelValle2012}. This provides an essentially perfect quantitative agreement with the
data without any processing such as deconvolution, provided, however, that one also includes the effect of the
anomalous moment term~$\mathcal{I}_1$ which bridges between the weak
and strong driving regimes. Indeed, $\Omega$ was not so low in the
experiment---in the interest of collecting enough signal in presence
of filtering---as to realize an ideal squeezed state to interfere with
the coherent fraction to produce the antibunching, but relied on a
distorted, skewed version of the squeezed state in its transition
towards the non-Gaussian, strong-driving regime where squeezing has
disappeared altogether. This term brings quantitative deviations which
are necessary to take into account to provide an exact match with the
data.  The unfiltered case, for instance, sees the vanishing-driving
two-photon statistics
$g^{(2)}(\tau)=\left[1-\exp(-\gamma_\sigma\tau/2)\right]^2$ turn into
\begin{equation}
  \label{eq:Fri17Apr2020144002BST}
  g^{(2)}(\tau) = 1 -e^{-3\gamma_\sigma \tau/4} \left[
    \cosh\left( \frac{R\tau}{4} \right) + \frac{3\gamma_\sigma}{R}
    \sinh \left( \frac{R\tau}{4} \right) \right]\,,
\end{equation}
at non-negligible driving, with~$R=\sqrt{\gamma_\sigma^2-64\Omega^2}$.  Using this and
numerically-exact filtered counterparts, with a global fitting that
only varies the filters widths and globally optimises the driving
strength~$\Omega$ and the two-level's decay
rate~$\gamma_\sigma=\SI{900}{\mega\hertz}$
(cf.~Table~\ref{tab:Fri1May111809CEST2020}), we obtain the solid lines
shown in Fig.~\ref{fig:g2}, providing an excellent
quantitative agreement with highly constrained fitting
parameters. From this data, one can extract the zero-delay coincidence
and compare it to the theory, i.e., both
Eq.~(\ref{eq:Fri17Apr143329CEST2020}), shown in dashed Red in
Fig.~\ref{fig:theory}, or to the finite~$\Omega$
counterpart that skews the squeezing, and whose expression is too bulky to be written
here~\cite{LopezCarreno2016}, but is given in the
Supplementary Material, Eq.~(\ref{eq:Fri1May105704CEST2020}). This also
yields an excellent agreement with the experimental data, which
confirms that filtering spoils antibunching according to the scenario
we have explained of perturbing the interference of the squeezed
fluctuations with the coherent signal, and that the experiment is
clean and fundamental enough to be reproduced exactly by including
non-vanishing driving features, without any further signal analysis or
data processing.

\begin{table}[t]
  \begin{ruledtabular}
    \begin{tabular}{c|c|c|c|c|c|c|c}
      Parameter & $\gamma_\sigma$ & $\Omega$ & $\Gamma_1$ & $\Gamma_2$
      & $\Gamma_3$ & $\Gamma_4$ & $\Gamma_5$ \\
      \hline
      Fitting (MHz) & 900 & 225 & 11403 & 1324 & 709 & 535 & 28 \\
      \hline
      \makecell{Data (MHz) \\ (Error)} & \makecell{890 \\ (60)} & \makecell{198 \\ (7)} & \makecell{19000 \\ (500)} & \makecell{1550 \\ (320)} & \makecell{780 \\ (160)} & \makecell{390 \\ (80)} & \makecell{28 \\ (6)}
    \end{tabular}
  \end{ruledtabular}
  \caption{Summary of the parameters used to fit the experimental
    data. The filters data are taken from the fabricant's data sheet,
    but are known to be typically measured in excess of their
    specified value.}
  \label{tab:Fri1May111809CEST2020}
\end{table}

In summary, we have shown that the emission from a two-level quantum system driven in the Rayleigh regime does not simultaneously yield subnatural linewidth and single-photon characteristics. When keeping only the subnatural linewidth part of the spectrum by frequency filtering, we do not observe antibunching in our second-order intensity correlation measurement. The narrower the spectral filtering, i.e., the fewer incoherently scattered photons we detect, the weaker the antibunching dip, which ultimately results in Poissonian photon statistics.
These results that disclose a perfect agreement with a fundamental
theory of time and frequency resolved photon correlations, with no
post-processing of the raw experimental data, are only the first step
towards a full exploitation of its consequences. In particular, since
the interference involves a coherent field, it is technically possible
to restore it fully in presence of filtering or, which is equivalent,
detection, simply by introducing externally the coherent fraction that is
missing or, in this case, in excess. This is done by
destructive interferences of the coherent signal, without perturbing
the quantum fluctuations. As a result, one should indeed obtain a
subnatural, laser-sharp, photon emission that is also perfectly
antibunched~\cite{LopezCarreno2018}. There are still other interesting
features in this regime, such as a plateau in the time-resolved photon
correlations. Such considerable improvements are in the wake of our
present findings.

Note added after proof: During the submission/preparation of the manuscript we became aware of a similar work~\cite{Phillips2020}.

\begin{acknowledgments}
This project has received funding from the European Union's Horizon 2020 research and innovation program under grant agreement No. 820423 (S2QUIP), the European Research Council (ERC) under the European Union’s Horizon 2020 Research and Innovation Programme (SPQRel, grant agreement no. 679183), Austrian Science Fund (FWF): P 29603, P 30459, the Linz Institute of Technology (LIT) and the LIT Lab for secure and correct systems, supported by the State of Upper Austria, the German Federal Ministry of Education and Research via the funding program Photonics Research Germany (contract number 13N14846), Q.Com (Project No. 16KIS0110) and Q.Link.X (16KIS 0874), the DFG via Project (SQAM) F1947/4-1, the Nanosystem Initiative Munich, the MCQST,  the  Knut  and  Alice Wallenberg  Foundation  grant  ”Quantum  Sensors”, the Swedish Research Council (VR) through the VR grant for international recruitment of leading researchers (Ref: 2013-7152), and Linn\ae{}us Excellence Center ADOPT.
K.M. acknowledges support from the Bavarian Academy of Sciences and Humanities. K.D.J. acknowledges funding from the Swedish Research Council (VR) via the starting Grant HyQRep (Ref 2018-04812) and The Göran Gustafsson Foundation (SweTeQ).
A.R. acknowledges fruitful discussions with Y. Huo, G. Weihs, R. Keil and S. Portalupi. 
\end{acknowledgments}

\bibliography{Rayleigh.bib}

\vfill\eject
\break\phantom{a}
\vfill\eject
\onecolumngrid
\section{The origin of antibunching in resonance fluorescence \\ Supplementary Material}

The theoretical description of a coherently driven quantum dot
modelled as a two-level system is straightforward with the formalism
of open quantum systems, e.g., writing the master equation (we
take~$\hbar=1$)
\begin{equation}
  \label{eq:Fri17Apr2020140226BST}
  \partial_t \rho = i[\rho,H_\sigma] + (\gamma_\sigma/2)\mathcal{L}_\sigma \rho\,,
\end{equation}
where~$H_\sigma=\Delta_\sigma\ud{\sigma}\sigma+\Omega (\ud{\sigma} +
\sigma)$ is the Hamiltonian of the quantum dot, with the two-level
system annihilation operator~$\sigma$ driven by a laser with
($c$-number) intensity~$\Omega$ with a detuning $\Delta_\sigma$, which
is zero in the conditions of our experiment (resonance). The
spontaneous decay of the quantum dot is modeled by the rightmost term
in Eq.~(\ref{eq:Fri17Apr2020140226BST}), where~$\gamma_\sigma$ is the
decay rate
and~$\mathcal{L}_c = (2c\rho \ud{c} - \rho\ud{c}c -
\ud{c}c\rho)$. Standard techniques yield the steady-state of
Eq.~(\ref{eq:Fri17Apr2020140226BST}) which is
\begin{equation}
\label{eq:rho2LS}
\rho = \left(1- n_\sigma \right) \ket{0} \bra{0} + \av{\sigma} \ket{0} \bra{1} + \av{\sigma}^{*} \ket{1} \bra{0} + n_\sigma  \ket{1} \bra{1} \,,
\end{equation}
where the total population $n_\sigma$ and mean field $\av{\sigma}$ are
\begin{equation}
  \label{eq:Tue19May100100CEST2020}
  n_\sigma = \frac{4 \Omega^2}{\gamma_\sigma^2 + 4 \Delta_\sigma^2 + 8 \Omega^2} \quad 
  \text{and} \quad
  \av{\sigma} = \frac{2 i \Omega \left(\gamma_\sigma - 2 i \Delta_\sigma \right)}{\gamma_\sigma^2 + 4 \Delta_\sigma^2 + 8 \Omega^2}\,.
\end{equation}
Equations~(\ref{eq:rho2LS}--\ref{eq:Tue19May100100CEST2020}) convey
well how $n_\sigma$ relates to the incoherent relaxation in the sense
of the two-level system being excited or spontaneously emitting
(diagonal elements of the density matrix) and how
$\langle\sigma\rangle$ relates to a coherent scattering connecting the
ground and excited states (off-diagonal elements). The
case~$\Delta_\sigma=0$ recovers Eqs.~(\ref{eq:Thu30Apr194924CEST2020})
of the text and their relative ratio as a function of pumping
power~$\Omega$ is shown in Fig.~\ref{fig:FPI}(c). A
useful and standard representation of the density matrix~$\rho$ is the
Wigner quasiprobability distribution
\begin{equation}
\label{eq:Tue19May100758CEST2020}
\tilde W(x,p) \equiv \frac{1}{\pi} \int_{-\infty}^{\infty} \bra{x+y} \rho \ket{x-y} e^{-2 i p y } dy \,,
\end{equation}
where $x$ and $p$ represent two conjugate observables which, in our
case, are proportional to the quantities of eventual interest, namely
the orthogonal set of (averaged) field quadratures $X$ and $Y$ (that
is, `position' is $x = \sqrt{2} X$ while `momentum' is
$p = \sqrt{2} Y$). Consequently, $\ket{x}$ is the eigenstate that
corresponds to the `position' operator . By substituting
Eq.~\eqref{eq:rho2LS} into~\eqref{eq:Tue19May100758CEST2020}, one gets
\begin{equation}
\tilde W_\sigma (x,p) = \left(1- n_\sigma \right) \tilde W_{00} + \av{\sigma}\tilde W_{01} + \av{\sigma}^{*} \tilde W_{10} + n_\sigma\tilde W_{11} \,,
\end{equation}
where $\tilde W_{mn}$ are the Wigner representations of the Fock state
matrix elements $\ket{m} \bra{n} $:
\begin{equation}
\tilde W_{mn} = \frac{1}{\pi} \int_{-\infty}^{\infty} \psi_{m}^{*} (x+y) \psi_{n} (x-y) e^{-2 i p y } dy \,.
\end{equation}
Within the integrand, the `position' representation of the
number-states $\ket{m}$ is:
\begin{equation}
\psi_m (x) = \bra{x} \ket{m} = \sqrt{\frac{1}{2^m m! \sqrt{\pi}}} e^{-x^2/2} H_m (x) \,,
\end{equation}
where $H_m (x)$ are the Hermite polynomials. After integration, we
obtain the next expressions:
\begin{subequations}
\begin{align}
  \tilde W_{00} & = \frac{1}{\pi} \, e^{-\left(x^2 + p^2 \right)} \,,\\
  \tilde W_{01} = \tilde W_{10}^{*}  & = \frac{\sqrt{2}}{\pi} \left(x + i p \right) e^{-\left(x^2 + p^2 \right)}  \,, \\
  \tilde W_{11} & = \frac{1}{\pi} \left(2 x^2 + 2 p^2 - 1 \right) e^{-\left(x^2 + p^2 \right)} \,.
\end{align}
\end{subequations}
We finally express the Wigner distribution in terms of the quadratures
of interest by changing $x \rightarrow \sqrt{2} X$ and
$p \rightarrow \sqrt{2} Y$ so the Wigner distribution reads:
\begin{equation}
   W_\sigma \left( X, Y \right) = 2 \,\tilde W_\sigma \left( \sqrt{2} X, \sqrt{2} Y \right) \,.
\end{equation}
The factor~2 is to preserve the normalization condition for the Wigner
distribution. The representation $W_\sigma$ is for the full
state~$\sigma$. It is simply related to~$W_\varsigma$ the
representation of the fluctuations by a mere translation in
phase-space:
\begin{equation}
  \label{eq:Tue19May105509CEST2020}
  W_\varsigma (X,Y) = W_\sigma \big(X - \Re\langle\sigma\rangle, Y - \Im\langle\sigma\rangle\big)
\end{equation}
since $\sigma$ and~$\varsigma$ are themselves related by the addition
of a coherent state
$\sigma=\varsigma+\langle\sigma\rangle$. Therefore, the shape of the
distribution is the same for both~$\sigma$
and~$\varsigma$. Furthermore, in the Rayleigh regime of weak-driving,
although~$\langle\sigma\rangle$ dominates over~$n_\sigma$ according to
Eq.~(\ref{eq:Tue19May100100CEST2020}), it is so small as to produce a
negligible displacement of the Gaussian cloud, itself in a thermal
squeezed state, which in appearance looks like a thermal state, since squeezing is too small compared to the thermal component to be seen with the naked eye. It is, however, essential to produce antibunching as a thermal and coherent admixture can only produce two-photon statistics between 1 and 2. In the
Fock regime of strong-driving, $\langle\sigma\rangle=0$ and
$W_\sigma=W_\varsigma$ exactly. In the intermediate case, the
bean-shaped skewed Wigner distribution is translated downward as the
result of sizable~$\langle\sigma\rangle$.

Detection and/or frequency filtering can be modelled with the
formalism of frequency-filtered and time-resolved $n$-photon
correlations~\cite{DelValle2012}, where the correlations of the
filtered light are obtained as the quantum averages of a ``sensor''
taken in the limit of its vanishing coupling to the emitter, otherwise
simply upgrading the Hamiltonian to
\begin{equation}
  \label{eq:Fri17Apr2020142351BST}
  H= H_\sigma + \epsilon(\ud{a}\sigma + \ud{\sigma}a)\,,
\end{equation}
where~$a$ is the annihilation operator of an harmonic oscillator that
models the sensor, which is $\epsilon$-coupled to the emitter, and the
master equation~(\ref{eq:Fri17Apr2020140226BST}) gets an additional
Lindblad term~$(\Gamma/2) \mathcal{L}_a\rho$, which describes the
bandwidth of the sensor and can be interpreted as the linewidth of an
interference (Lorentzian-shaped) filter. In the steady-state and for a
generic operator~$c$, the second-order correlations are defined
as~\cite{Glauber1963}
\begin{equation}
  \label{eq:Fri17Apr2020143710BST}
  g^{(2)}(\tau) = \frac{\mean{\ud{c}(\ud{c}c)(\tau)c}}{\mean{\ud{c}c}^2}\,,
\end{equation}
and such quantities can be obtained for the sensor~$a$ according to
the standard techniques, thereby providing
easily quantities of direct and high experimental interest, such as
those discussed in the main text (in particular~$g^{(2)}(\tau)$ as
shown in Fig.~3), without recourse to processing of the raw data.  One
quantity of great significance, and that can be obtained in this way,
is the zero-delay two-photon correlator~$g^{(2)}(0)$ at arbitrary
driving~$\Omega$, which can be found by considering the cascaded
excitation of an harmonic oscillator by the coherent single-photon
source~\cite{LopezCarreno2016}.
Defining~$\digamma_{kl}\equiv k\Gamma+l\gamma_\sigma$ for
integers~$k$, $l$ (e.g., $\digamma_{11}=\Gamma+\gamma_\sigma$), the
two-photon coincidence correlation function is found
as~\cite{LopezCarreno2016}:
\begin{equation}
  \label{eq:Fri1May105704CEST2020}
  g^{(2)}(\tau=0)=\frac{\digamma_{11}\left(\gamma_\sigma ^2+4 \Omega ^2\right) \left(\digamma_{11}\digamma_{12}+8 \Omega ^2\right) 
    \left(48 \Gamma ^2 \Omega ^4 \digamma_{21}+4 \Gamma  \Omega ^2 \digamma_{31} \left(17 \Gamma ^3+29 \Gamma ^2 \gamma_\sigma +18 \Gamma  \gamma_\sigma ^2+4 \gamma_\sigma ^3\right)
      +\digamma_{11}\digamma_{21}^2\digamma_{31}^2 \digamma_{12}\digamma_{32}\right)}
  {\digamma_{21}\digamma_{31}\left(\digamma_{11}\digamma_{21}+4 \Omega ^2\right) \left(\digamma_{31}\digamma_{32}+8 \Omega ^2\right) 
    \left(\digamma_{12}\digamma_{11}^2 +4 \Gamma  \Omega ^2\right)^2}\,.
\end{equation}
This exact analytical expression describes with excellent accuracy the
loss of antibunching observed in our experiment due to frequency
filtering and taking into account anomalous correlations in the
squeezing of the incoherent signal.  This shows among other things
that in the intermediate driving regime, $g^{(2)}(0)$ can be larger
than~1, i.e., the coherently-driven two-level system can emit bunched
filtered photons, in direct opposition to their previously assumed
antibunched character.

\end{document}